\documentclass[aps,prb,twocolumn,preprintnumbers,amsmath,amssymb,superscriptaddress]{revtex4}%

\usepackage{graphicx}
\usepackage{dcolumn}
\usepackage{amsmath}
\usepackage{color}
\usepackage{hyperref}
\usepackage{bbold}

\newcommand{\RN}[1]{\textup{\uppercase\expandafter{\romannumeral#1}}}%

\begin{document}

\title{Pressure-enhanced splitting of density wave transitions in La$_3$Ni$_2$O$_{7-\delta}$}

\author{Rustem Khasanov}
 \email{rustem.khasanov@psi.ch}
 \affiliation{Laboratory for Muon Spin Spectroscopy, Paul Scherrer Institut, CH-5232 Villigen PSI, Switzerland}

\author{Thomas J. Hicken}
 \affiliation{Laboratory for Muon Spin Spectroscopy, Paul Scherrer Institut, CH-5232 Villigen PSI, Switzerland}

\author{Dariusz J. Gawryluk}
 \affiliation{Laboratory for Multiscale Materials Experiments, Paul Scherrer Institut, CH-5232 Villigen PSI, Switzerland}

\author{Vahid Sazgari}
 \affiliation{Laboratory for Muon Spin Spectroscopy, Paul Scherrer Institut, CH-5232 Villigen PSI, Switzerland}

\author{Igor Plokhikh}
\affiliation{Laboratory for Multiscale Materials Experiments, Paul Scherrer Institut, CH-5232 Villigen PSI, Switzerland}

\author{Lo\"{i}c Pierre Sorel}
 \affiliation{Laboratory for Multiscale Materials Experiments, Paul Scherrer Institut, CH-5232 Villigen PSI, Switzerland}

\author{Marek Bartkowiak}
 \affiliation{Laboratory for Neutron and Muon Instrumentation, Paul Scherrer Institute, Villigen PSI CH-5232, Switzerland}

\author{Steffen B\"{o}tzel}
 \affiliation{Theoretische Physik III, Ruhr-Universit\"{a}t Bochum, D-44780 Bochum, Germany}

\author{Frank Lechermann}
 \affiliation{Theoretische Physik III, Ruhr-Universit\"{a}t Bochum, D-44780 Bochum, Germany}

\author{Ilya M. Eremin}
 \affiliation{Theoretische Physik III, Ruhr-Universit\"{a}t Bochum, D-44780 Bochum, Germany}

\author{Hubertus Luetkens}
 \affiliation{Laboratory for Muon Spin Spectroscopy, Paul Scherrer Institut, CH-5232 Villigen PSI, Switzerland}

\author{Zurab Guguchia}
 \affiliation{Laboratory for Muon Spin Spectroscopy, Paul Scherrer Institut, CH-5232 Villigen PSI, Switzerland}

\begin{abstract}
The observation of superconductivity in La$_3$Ni$_2$O$_{7-\delta}$ under pressure, following the suppression of a high-temperature density wave state, has attracted considerable attention. The nature of this density wave order was not clearly identified. Here, we probe the magnetic response of the zero-pressure phase of La$_3$Ni$_2$O$_{7-\delta}$ as hydrostatic pressure is applied and find that the apparent single density wave transition at zero applied pressure splits into two. The comparison of our muon-spin rotation and relaxation experiments with dipole-field numerical analysis reveals the magnetic structure's compatibility with a stripe-type arrangement of Ni moments, characterized by alternating lines of magnetic moments and nonmagnetic stripes at ambient pressure. When pressure is applied, the magnetic ordering temperature increases, while the unidentified density wave transition temperature falls. Our findings reveal that the ground state of the La$_3$Ni$_2$O$_{7-\delta}$ system is characterized by the coexistence of two distinct orders -- a magnetically ordered spin density wave and a lower-temperature ordering that is most likely a charge density wave -- with a notable pressure-enhanced separation between them.

\end{abstract}

\maketitle

\section{Introduction}
\begin{figure}[htb]
\includegraphics[width=0.85\linewidth]{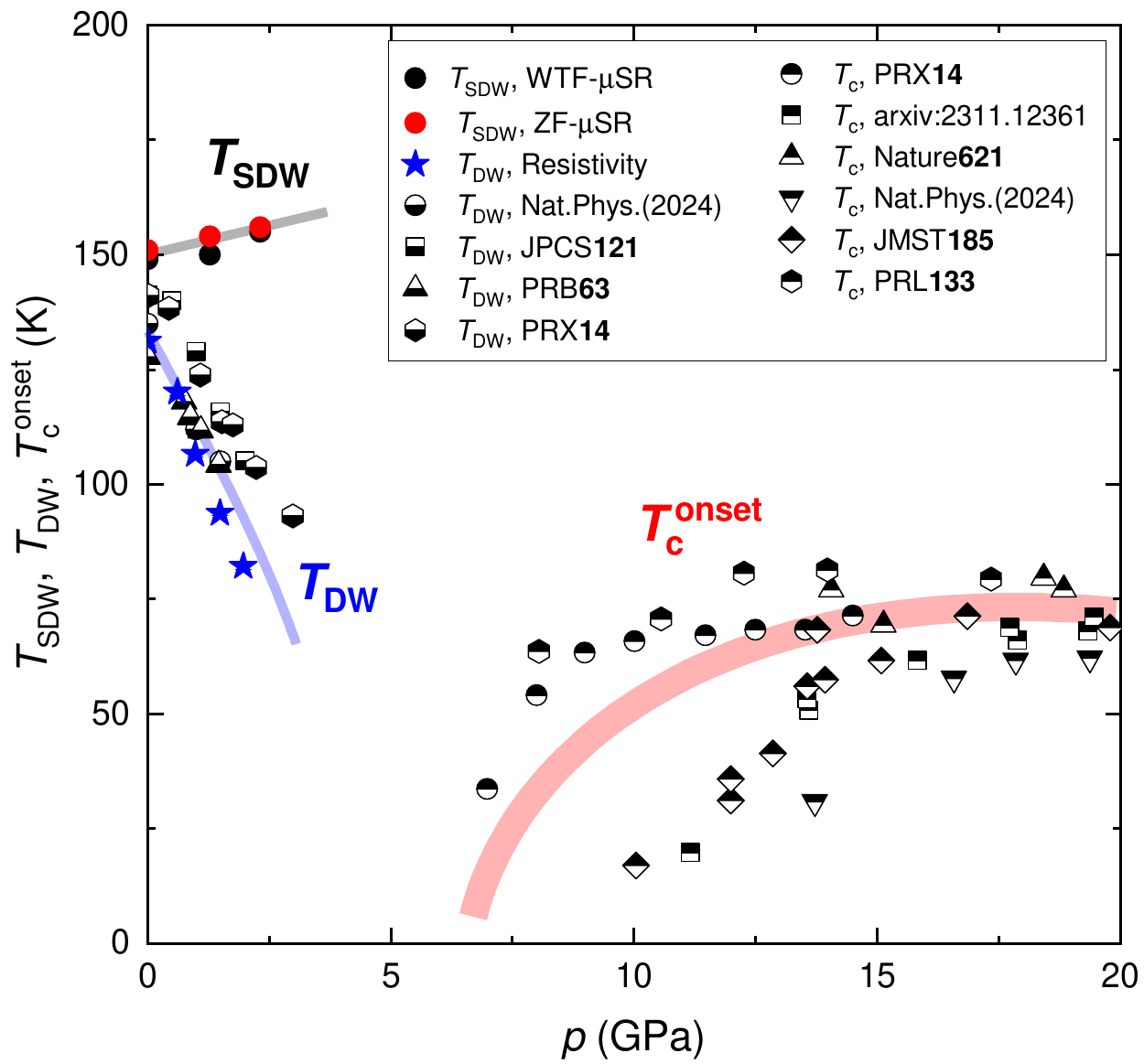}
\caption{{\bf The conjectured pressure-temperature ($p-T$) phase diagram of La$_3$Ni$_2$O$_{7-\delta}$.} The solid symbols represent pressure dependence of the SDW ordering temperature ($T_{\rm SDW}$) as obtained in ZF and WTF-$\mu$SR experiments (closed red and black circles) and the DW ordering temperature ($T_{\rm DW}$) obtained in resistivity studies (blue stars) in the present work. The half-down filled symbols are $T_{\rm DW}$ vs. $p$ data from Refs.~\onlinecite{Wang_Arxiv_2023, Zhang_arxiv_2023, Hosoya_JPCS_2008, Wu_PRB_2023}. The half-up filled symbols represent the pressure dependence of  the onset of the superconducting transition temperature  $T_{\rm c}^{\rm onset}$ from Refs.~\onlinecite{Sun_Nature_2023, Zhang_JMST_2024, Hou_CPL_2023, Wang_Arxiv_2023, Zhang_arxiv_2023, Puphal_arxiv_2023}. The lines are guides for the eye. }
 \label{fig:Phase-Diagram}
\end{figure}

The discovery of superconductivity in La$_{3}$Ni$_{2}$O$_{7-\delta}$  under pressure ($p$)  has attracted significant attention.\cite{Sun_Nature_2023, Zhang_JMST_2024, Hou_CPL_2023, Wang_Arxiv_2023, Wang_review_arxiv_2024}
This interest is accentuated by the material's critical temperature $T_{\rm c}\simeq 80$~K, which is notably above the boiling point of liquid nitrogen, positioning La$_{3}$Ni$_{2}$O$_{7-\delta}$ within the category of high-temperature superconducting materials. The recent study of Wang et al.\cite{Wang_Arxiv_2023} suggests that superconductivity in La$_{3}$Ni$_{2}$O$_{7-\delta}$ emerges upon the suppression of a competing density wave (DW) order. It was found, in particular, that the DW-like anomaly in resistivity, which sets in at ambient pressure around $T_{\rm DW} \sim 140$~K, is progressively suppressed with increasing $p$, while the onset of a superconducting transition emerges at $p\sim7$~GPa, see Fig.~\ref{fig:Phase-Diagram}. Similar DW anomalies were detected at ambient pressure in specific heat and magnetization experiments,\cite{Wu_PRB_2001, Liu_ScChinaMechAstr_2023} as well as by means of NMR. \cite{Fukamachi_JPSJ_2001, Kakoi_Arxiv_2023} However, the origin of the competing DW order remains a matter of ongoing investigations.
Charge ordering in NiO$_2$ planes induced by oxygen orderings,\cite{Taniguchi_JPSJ_1995, Kobayashi_JPSJ_1996} or charge-density wave instabilities induced by one-dimensional Fermi surface nesting,\cite{Wu_PRB_2001, Seo_InChem_1966} were suggested to account for these anomalies. However there is no direct evidence for the charge ordering in La$_3$Ni$_2$O$_{7-\delta}$ reported up to now.
In a more recent work, Chen et al.\cite{Chen_arxiv_2023} have reported the observation of a spin density wave (SDW) type of order in La$_{3}$Ni$_{2}$O$_{7-\delta}$  by means of the  muon-spin rotation/relaxation ($\mu$SR) technique.  The ambient pressure zero-field $\mu$SR experiments reveal the presence of static long-range magnetic order with the transition temperature $T_{\rm N}\simeq 148$~K. The SDW ordering was further confirmed by the  resonant inelastic X-ray scattering,\cite{Chen_arxiv_2024} and NMR experiments.\cite{Dan_arxiv_2024}

Consequently, a question arises: whether the competing DW order is magnetic in origin, as suggested by the similar ambient pressure values of DW transition temperatures $T_{\rm DW} \sim 140-150$~K \cite{Wang_Arxiv_2023, Wu_PRB_2001, Liu_ScChinaMechAstr_2023, Fukamachi_JPSJ_2001, Kakoi_Arxiv_2023, Zhang_arxiv_2023, Hosoya_JPCS_2008, Wu_PRB_2023} and $T_{\rm SDW} \sim 150$~K,\cite{Dan_arxiv_2024, Chen_arxiv_2023} or is the unspecified DW different in origin and coexisting with the SDW? Notably, in the single-layer La$_2$NiO$_{4}$ and the  three-layer La$_4$Ni$_3$O$_{10}$ Ruddlesden-Popper lanthanum nickelates the coexistent spin and charge orders were detected.\cite{Tranquada_PRB_1995, Zhang_NatCom_2020} In order to address this question, in this work comprehensive $\mu$SR and resistivity experiments under hydrostatic pressure conditions were performed. The primary aim was to determine whether or not the SDW order follows the same pressure dependence as the DW order. At ambient pressure, our results corroborate the findings of Refs.~\onlinecite{Chen_arxiv_2023, Dan_arxiv_2024} detecting a spin density wave (SDW) type of magnetism with a transition temperature $T_{\rm N}\simeq 151$~K. The DW transition, corresponding to the local minimum on resistivity curve, sets 20~K lower at $T_{\rm DW}\simeq 131$~K.
Under increasing pressure, the magnetic ordering temperature was observed to rise at a rate ${\rm d}T_{\rm N}/{\rm d}p\simeq 2.8$~K/GPa, which is not only opposite in sign but also substantially smaller in magnitude compared to the ${\rm d}T_{\rm DW}/{\rm d}p\simeq-26$~K/GPa, see Fig.~\ref{fig:Phase-Diagram}. The comparison of our experimental data with magnetic dipole-field calculations also corroborates the notion of coexisting magnetic SDW and non-magnetic DW orders. Both these observations suggest that the competing DW order, which is suggested to compete with superconductivity in La$_{3}$Ni$_{2}$O$_{7-\delta}$, is, likely, non-magnetic in origin.
These results contribute to a deeper understanding of the intricate relationship between magnetism and superconductivity in this complex oxide, offering new insights into the nature of the competing phases in high-temperature superconductors.
\\

\section{Results}
\subsection{Ambient pressure $\mu$SR}

\begin{figure}[htb]
\includegraphics[width=1.0\linewidth]{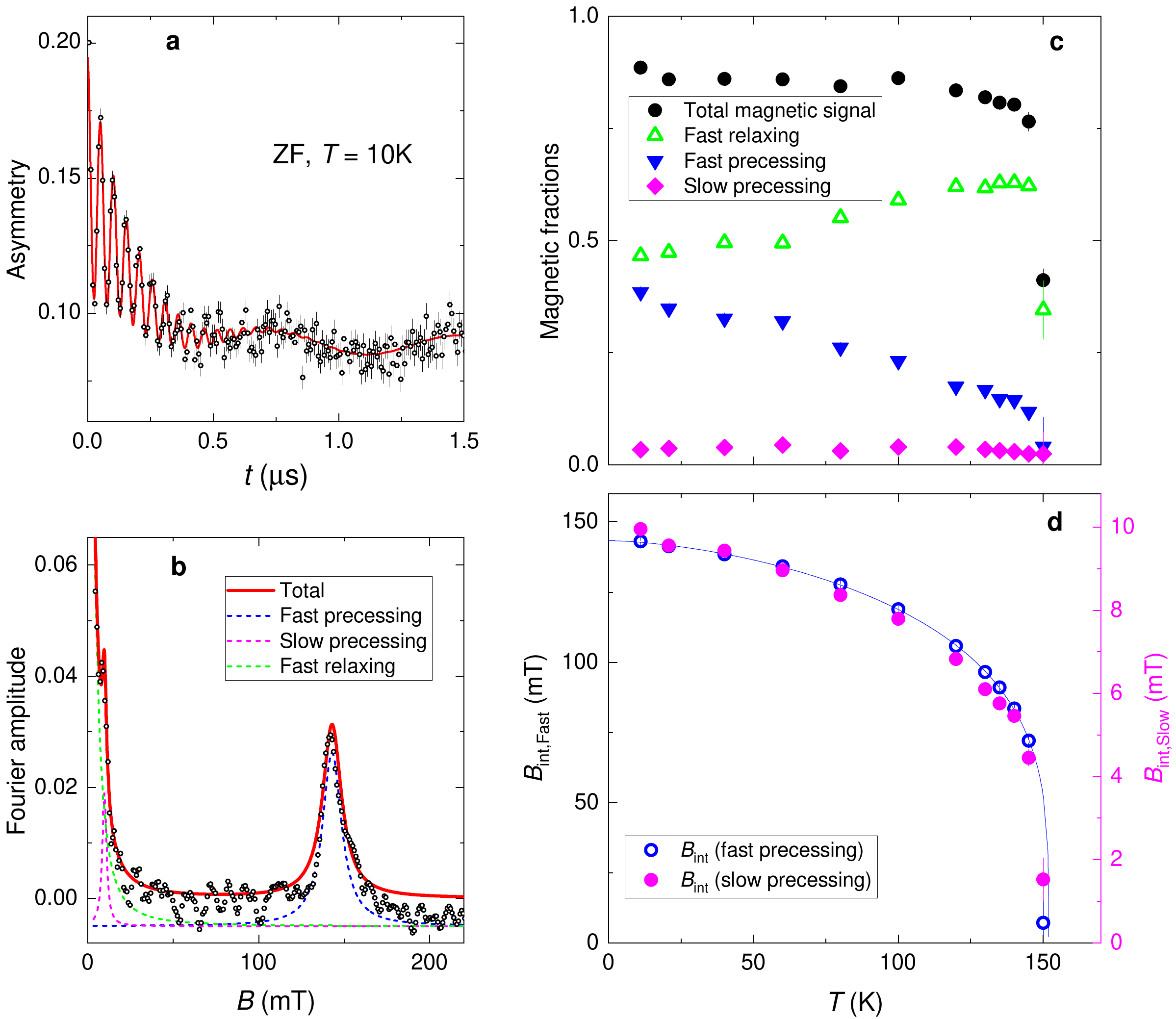}
\caption{{\bf The results of ambient pressure $\mu$SR experiments.} {\bf a} The zero-field $\mu$SR time spectra of the La$_3$Ni$_2$O$_{7-\delta}$ sample measured at $T=10$~K. The red line is fit of Eq.~\ref{eq:asymmetry} to the data. {\bf b} The Fourier transform of the data presented in panel {\bf a}. The dashed lines represent individual fit components. {\bf c} $T$ dependencies of the ZF-$\mu$SR signal fractions. {\bf d} $T$ dependencies of the internal field of the fast precessing and the slow precessing components. The solid line is the fit of the power law, Eq.~\ref{eq:power-law}, to the $B_{\rm int,Fast}(T)$ data. The displayed error bars for parameters obtained from $\mu$SR data correspond to one standard deviation from the $\chi^2$ fits.  }
 \label{fig:Ambient-pressure}
\end{figure}

The ZF-$\mu$SR response of La$_{3}$Ni$_{2}$O$_{7-\delta}$ measured at ambient pressure at $T=10$~K is presented Fig.~\ref{fig:Ambient-pressure}. Panels {\bf a} and {\bf b} correspond to the ZF asymmetry spectra, representing the time evolution of the muon-spin polarization, and the Fourier transformation, showing the distribution of the internal fields, respectively.

The data were analyzed by using Eqs.~\ref{eq:asymmetry} and \ref{eq:commensurate} described in the Method section.
The fit results reveal that the transversal part of Eq.~\ref{eq:commensurate} consists of three terms: the fast precessing, slow precessing and non precessing fast relaxing ones [see the notations in Fig.~\ref{fig:Ambient-pressure}~{\bf b}]. The corresponding internal fields ($B_{\rm int}$'s) at $T=10$~K are 0.15, 0.01, and 0.0~T. This is reminiscent of the low-$T$ ($T\lesssim 40$~K) ZF-$\mu$SR results of LaFeAsO (i.e., the parent compound of 1111 family of Fe-based superconductors) consisting of three magnetic components with the corresponding internal field values 0.17, 0.02, and 0.0~T.\cite{Klauss_PRL_2008} Note that
the fit with the simple cosine type of oscillating functions, Eq.~\ref{eq:commensurate}, suggests the consistency of our ZF-$\mu$SR dat with the commensurate magnetic order (see the Supplementary Note 4 and Supplementary Fig. 3).

The temperature evolution of the magnetic fractions is presented in Fig.~\ref{fig:Ambient-pressure}~{\bf c}. The analysis reveals that for $T\lesssim 150$~K, approximately $90$\% of the full asymmetry is assigned to the magnetic contribution. Considering that the remaining 10\% might be partially explained by the contribution of muons stopped outside of the sample ($3-5\%$, in line with the sample dimensions and GPS spectrometer characteristics, Ref.~\onlinecite{GPS_web-page}) as well as by muons stopped within the lanthanum silica apatite impurity phase (of the order of 8\%, see the Supplementary Note 1 and Supplementary Fig. 1), this suggests that magnetism in the La$_3$Ni$_2$O$_{7-\delta}$ sample studied here is representative for the bulk. The transition from the magnetic to the nonmagnetic state (from $f_{\rm m}\simeq 0.9$ to $f_{\rm m}=0$) is rather sharp, thus suggesting that magnetism in La$_3$Ni$_2$O$_{7-\delta}$ sets in homogeneously.

Figure~\ref{fig:Ambient-pressure}~{\bf c} shows that the volume fraction of the slow oscillating component remains nearly temperature-independent, while fractions of the fast relaxing and fast oscillating components show opposite trends.  With increasing temperature, the fast relaxing signal develops at the expense of the fast oscillating component [Fig.~\ref{fig:Ambient-pressure}~{\bf c}].
A possible explanation is that the fast relaxing component originates from muons that stop in highly magnetically disordered regions, hence experiencing a large distribution of fields and dephasing quickly. Conversely, the fast precessing component is from muons that sit in the same crystallographic site, but in regions of well-ordered magnetism. The change in amplitude suggest that as the temperature increases, the volume fraction of disordered-type sites goes up (hence the relaxing fraction increases) at the expense of the well-ordered regions (hence the oscillating fraction drops). This could happen if, for example, the correlation length decreases with increasing temperature, meaning the ordered regions shrink and the disordered ones grow.

\begin{figure}[htb]
	\centering
	\includegraphics[width=0.9\linewidth]{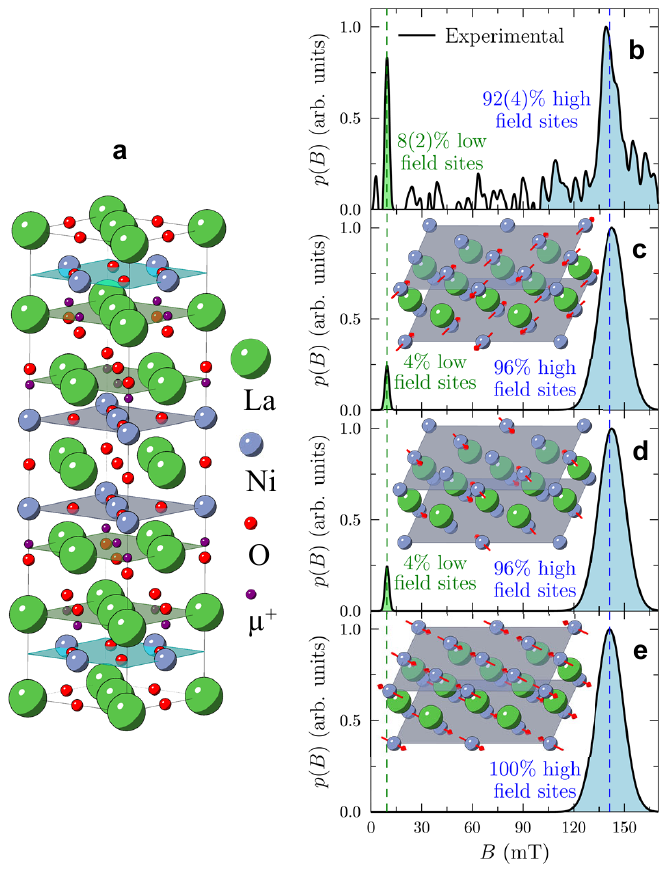}
    \includegraphics[width=1.0\linewidth]{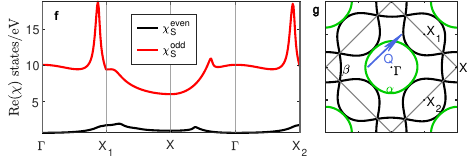}
	\caption{{\bf The candidate magnetic structures of La$_3$Ni$_2$O$_{7-\delta}$.} {\bf a} The muon stopping sites as calculated with DFT$+\mu$. Despite all shown sites being crystallographically equivalent, for some magnetic structures they are not magnetically equivalent. The Ni and La planes are highlighted in different colours. {\bf b} The magnetic field distribution $p(B)$ from Fig.~\ref{fig:Ambient-pressure}~{\bf b}, but with the zero-field peak removed, highlights two internal fields seen by muons. {\bf c}--{\bf e} Simulations of $p(B)$'s for different magnetic configurations. The magnetic unit cells are shown as insets. O atoms are hidden for clarity. {\bf f} The spin susceptibility for the tight-binding parameterization from Ref.~\onlinecite{Wang_arxiv_2024}. {\bf g} The Fermi surface and the dominant scattering vector $\mathbf{Q}_\text{SDW} \approx X_1$.}
	\label{fig:muonSite_dipoleField}
\end{figure}

The internal magnetic field (i.e., the magnetic order parameter) decreases with increasing temperature and vanishes at $T_{\rm N}$, Fig.~\ref{fig:Ambient-pressure}~{\bf d}.
The temperature behaviour of both the fast precessing ($B_{\rm int,Fast}$)  and the slow precessing ($B_{\rm int,Slow}$) internal fields remain nearly the same [Fig.~\ref{fig:Ambient-pressure}~{\bf d}], suggesting the presence of at least two nonequivalent muon stopping sites within the magnetic unit cell of La$_3$Ni$_2$O$_{7-\delta}$.
The solid line in Fig.~\ref{fig:Ambient-pressure}~{\bf b} represents a fit of a phenomenological power-law function expressed as:\cite{Pratt_JPCM__2007}
\begin{equation}
B_{\rm int}(T)= B_{\rm int}(0) [1-(T/T_{\rm N})^\alpha]^\beta
 \label{eq:power-law}
\end{equation}
to $B_{\rm int}(T)$ of the fast oscillating component. The fit results in the magnetic transition temperature $T_{\rm N}=151.5(1.2)$~K, and exponents $\alpha=1.58(9)$ and $\beta=0.26(2)$. The obtained magnetic ordering temperature is in agreement with that of recent resistivity,\cite{Wang_Arxiv_2023, Liu_ScChinaMechAstr_2023} NMR,\cite{Kakoi_Arxiv_2023, Dan_arxiv_2024} and  $\mu$SR experiments.\cite{Chen_arxiv_2023} The exponent $\beta\simeq0.26$ suggests a second-order-type phase transition and is rather close to a critical exponent of $1/3$ expected for 3D magnetic systems.\cite{Campostrini_PRB_2002}
\\

\subsection{The candidate magnetic structure(s) of La$_3$Ni$_2$O$_{7-\delta}$}

Different magnetic ground states were proposed for  La$_3$Ni$_2$O$_7$,\cite{Chen_arxiv_2024, Xie_arxiv_2024, Dan_arxiv_2024} however no technique has thus far provided an unambiguous determination.
To explore the feasibility of candidate magnetic structures the dipole-field at muon stopping sites were calculated.
The proposed structures focus on Ni moment arrangement within the $ab-$plane, which leaves a large number of possible configurations both due to possible different $c-$axis stacking patterns, where weak intralayer coupling was suggested,\cite{Chen_arxiv_2024} and by changing the moment direction.
Varying either of these parameters does not qualitatively change the results, hence the best match to experiment for different $ab-$plane configuration are presented in Fig.~\ref{fig:muonSite_dipoleField}.
One important result is that some of the Ni sites lack magnetic moment, otherwise the slow precession component (corresponding to a low magnetic field at the muon site) is not observed.
This suggests that La$_3$Ni$_2$O$_7$  must host charge- and spin density wave, analogous to a stripe-type order in cuprate high-temperature superconductors.

There are two proposed arrangements of the missing moments, the first [Fig.~\ref{fig:muonSite_dipoleField}~{\bf c}] with two adjacent lines of vacancies in the [110] direction, and the second [Fig.~\ref{fig:muonSite_dipoleField}~{\bf d}] with only one line separated by lines of moments. Experimentally $B_{\rm int,Fast}/B_{\rm int,Slow}\simeq15$ [Fig.~\ref{fig:Ambient-pressure}~{\bf c}], which can be obtained in for either vacancy structure by varying the $c-$axis stacking and moment direction.
The moment carried by Ni ($m_{\rm Ni}$) might be estimated from the magnitude of the field at the muon site (Supplementary Note 8 and Supplementary Table II).
If the moment points in the $ab-$plane, then $m_{\rm Ni}=0.48$--$0.67~\mu_{\rm B}$, in good agreement with the fluctuating moment of $m_{\rm Ni}=0.55~\mu_{\rm B}$ predicted in Ref.~\onlinecite{shilenko2023correlated}. Conversely a lower value is required if the moment points parallel to the $c-$axis, $m_{\rm Ni} = 0.28$--$0.31~\mu_{\rm B}$.
In addition, our calculations exclude all magnetic structure models without zero moment lines, {i.e., models with magnetic moments on all Ni sites, as these consistently would not exhibit low field muon sites in contrast to our experimental observation, see Figs.~\ref{fig:muonSite_dipoleField}~{\bf c} and {\bf e}.

A long-range commensurate magnetic order with the ordered moments ranging from 0.3 to 0.7 ~$\mu_{\rm B}$ should be detected by neutron experiments. The fact that neutrons do not see the magnetic order\cite{Xie_arxiv_2024} might be related to the difference in correlation length between $\mu$SR and neutron techniques. Muons are known to require a shorter coherence length in order to obtain an oscillatory signal.\cite{Yaouanc_book_2011, Dalmas_JPCS_2014}  An indirect confirmation of a short magnetic coherence length in La$_3$Ni$_2$O$_7$ might be the presence of the non-oscillating fast relaxing component, which increases as it approaches $T_{\rm N}$ [see Figs.~\ref{fig:Ambient-pressure}~{\bf b} and {\bf c}]. It seems the short magnetic correlation length limits the amount of coherent magnetic fraction already at low temperatures, while the development of the fast relaxing signal at the expense of the fast oscillating component suggests further shortening of the coherence length as a function of temperature.

\begin{figure}[htb]
\includegraphics[width=1.0\linewidth]{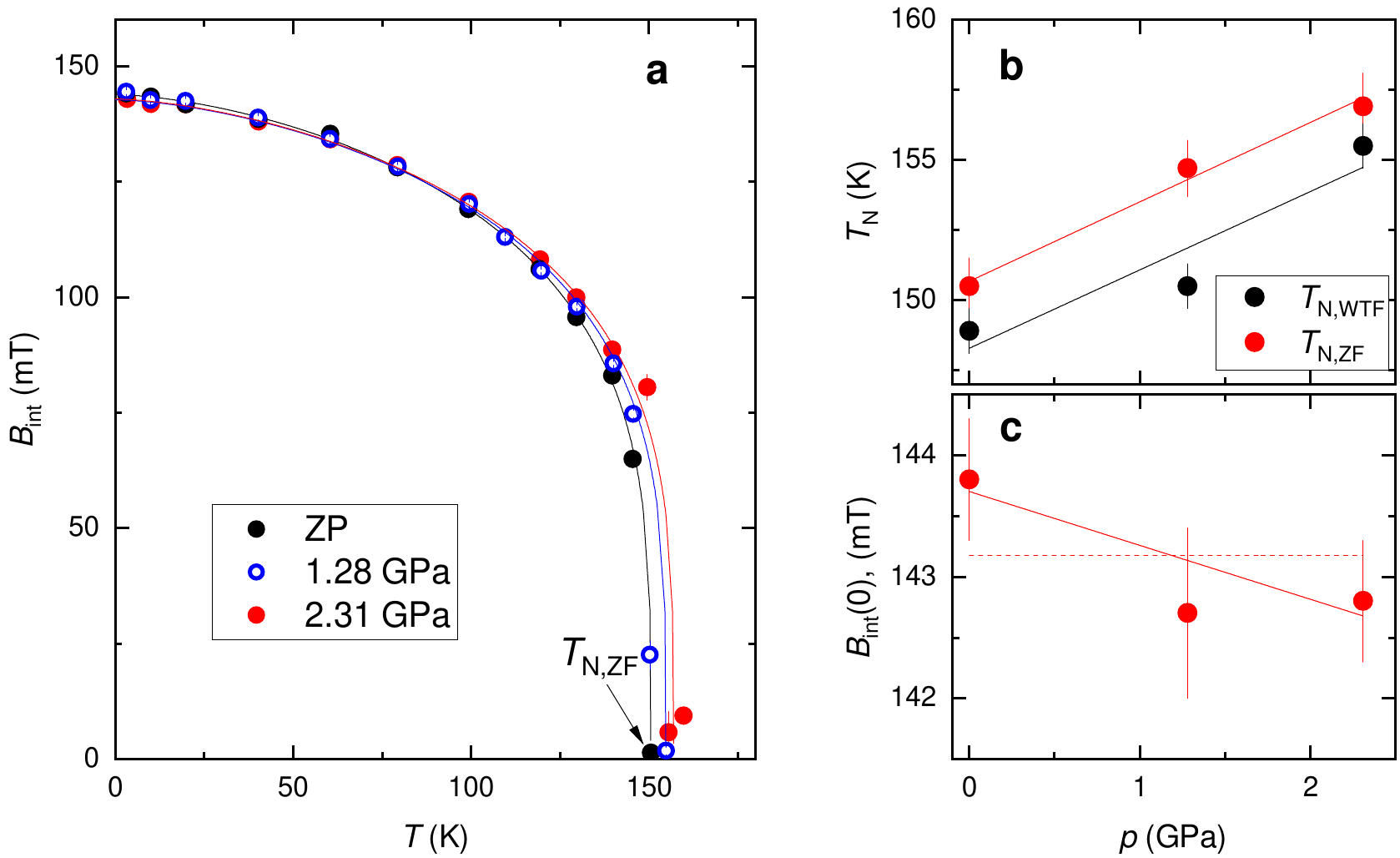}
\caption{{\bf Pressure evolution of SDW order.} {\bf a} Temperature dependencies of the internal field of the fast precessing component measured at pressures $p=0.0$, 1.28, and 2.31~GPa.  {\bf b} Pressure dependencies of the magnetic ordering temperatures as determined in WTF and ZF-$\mu$SR experiments. The solid lines are linear fits with $T_{\rm N,WTF}(p) = 148.3(1.5)+p\cdot2.8(4)$ and $T_{\rm N,ZF}= 150.7(5)+p\cdot2.8(3)$. {\bf c} The pressure dependence of the internal field of the fast precessing component. The solid and dashed lines are linear fits with $B_{\rm int}(0,p)= 143.7(3)-p\cdot0.44(22)$ and $B_{\rm int}(0,p)= 143.2(4)$, respectively. The error bars for individual data points correspond to one standard deviation from the $\chi^2$ fits.}
 \label{fig:High-pressure}
\end{figure}

To proceed further with possible magnetic structures, the spin susceptibility at ambient pressure in the paramagnetic state was calculated within the framework of a tight-binding parametrization from Ref.~\onlinecite{Wang_arxiv_2024}.  Note that in bilayer systems, like La$_3$N$_2$O$_7$, the spin susceptibility can be decomposed into the even ($\chi_{\rm S}^{\rm even}$) and odd ($\chi_{\rm S}^{\rm odd}$) channels following the scheme outlined in Ref.~\onlinecite{Botzel2024}. The resulting $\chi_{\rm S}^{\rm odd}$ and $\chi_{\rm S}^{\rm even}$ components of the spin susceptibility are presented in Fig.~\ref{fig:muonSite_dipoleField}~{\bf f}. The details of the calculations are provided in the Supplementary Note 9.
The dominant peak in odd susceptibility is connected to scattering between the bonding $\alpha$ and antibonding $\beta$ bands and it peaks near the wave vector ${\bf Q}_{SDW} \approx (\pi/2,\pi/2)=X_1$, which is visualized within the pseudo-tetragonal description\cite{Chen_arxiv_2024}  in Fig.~\ref{fig:muonSite_dipoleField}~{\bf g}. Peaks around $X_1=(\pi/2,\pi/2)$ and $X_2=(\pi/2,-\pi/2)$ are slightly different in magnitudes due to intrinsic orthorhombicity in the system, making instability along $X_1$ more preferable.
The peak at {\bf Q}$_{SDW}$ implies instability towards double stripe order within each layer and, since it occurs in the odd spin channel (i.e., for $q_z=\pi/d$, where $d$ is the thickness of the bilayer sandwich), the antiferromagnetic coupling between the stripes along $c-$direction. Note, that to project the magnetization into the real space and make comparison with our experimental finding, the magnetization on the lattice can be written as $\vec{M}_1({\bf R}_j)=\vec{\Delta}_{SDW} \cos \left({\bf Q}_{SDW} \cdot {\bf R}_j +\pi/4 \right)$ for Fig.~\ref{fig:muonSite_dipoleField}~{\bf e} and $\vec{M}_2({\bf R}_j)=\vec{\Delta}_{SDW} \cos \left({\bf Q}_{SDW} \cdot {\bf R}_j + \pi/2 \right)$ for Fig.~\ref{fig:muonSite_dipoleField}~{\bf d}, but requires inclusion of the higher harmonics for Fig.~\ref{fig:muonSite_dipoleField}~{\bf c} as $\vec{M}_3({\bf R}_j)= \vec{\Delta}_{SDW} \left[ \sqrt{2} \cos \left({\bf Q}_{SDW} \cdot {\bf R}_j + \pi/4 \right)+\cos \left(2{\bf Q}_{SDW} \cdot {\bf R}_j \right) \right]$.

\subsection{High-pressure $\mu$SR and resistivity experiments}

The $\mu$SR experiments under quasi-hydrostatic pressure conditions were performed at pressures $p=0.0$, 1.28, and 2.31~GPa. Two sets of experiments were conducted: the first with a weak (5~mT) magnetic field applied perpendicular to the initial muon spin polarization (WTF) and the second performed in zero applied field (ZF). The details of the ZF- and WTF-$\mu$SR data analysis procedure, as well the determination of the magnetic ordering temperature from WTF-$\mu$SR data ($T_{\rm N,WTF}$) are discussed in the Supplementary Notes 4 and 5.

\begin{figure}[htb]
\includegraphics[width=1.0\linewidth]{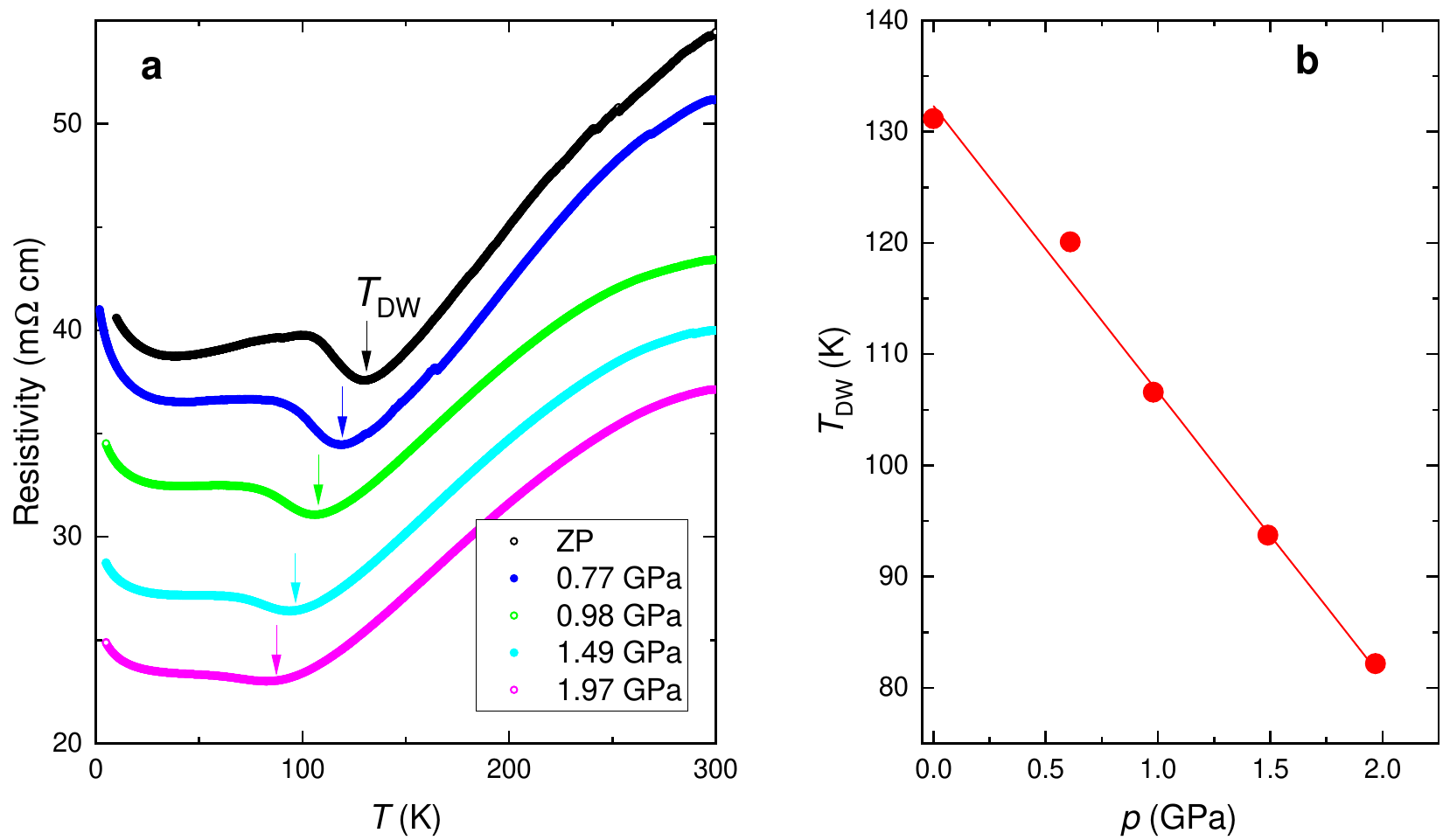}
\caption{{\bf Pressure evolution of DW order.} {\bf a} Temperature dependencies of resistivity measured at pressures $p=0.0$, 0.61, 0.98, 1.49 and 1.97~GPa. Arrows represent the DW transition temperature $T_{\rm DW}$. {\bf b} Dependence of $T_{\rm DW}$ on pressure. The solid line is the linear fit: $T_{\rm DW}(p)=132.3(1.6)-p\cdot25.7(1.2)$.}
 \label{fig:High-pressure_resisitivity}
\end{figure}

Temperature dependencies of the internal field of the fast precessing component ($B_{\rm int, Fast}$) measured at various pressures are presented in Fig.~\ref{fig:High-pressure}. The solid lines correspond to the fit of Eq.~\ref{eq:power-law} to $B_{\rm int}(T)$ data. To avoid the high sensitivity of the fit parameters [i.e., $B_{\rm int}(0)$ and $T_{\rm N}$] to the values of the exponents $\alpha$ and $\beta$, they were fixed at $\alpha=1.58$ and $\beta=0.26$ as obtained in ambient pressure studies [see the text above and Figs.~\ref{fig:Ambient-pressure}~{\bf d} and \ref{fig:High-pressure}~{\bf a}].
The pressure dependencies of fit parameters, namely the magnetic ordering temperatures $T_{\rm N,WTF}$,  $T_{\rm N,ZF}$, and the zero-temperature value of the internal field $B_{\rm int}(0)$, are plotted in Figs.~\ref{fig:High-pressure}~{\bf b} and {\bf c}. Both $T_{\rm N}$s, as determined in WTF- and ZF-$\mu$SR experiments, increase with increasing pressure. The linear fits result in equal slopes with ${\rm d} T_{\rm N}/ {\rm d}p=2.8(3)$~K/GPa. The internal field, which is proportional to the value of the ordered magnetic moment on Ni site [$B_{\rm int}(0)\propto m_{\rm Ni}$], shows different tendency. The ordered moment is nearly pressure independent, or may even slightly decrease with increasing pressure as: \\ ${\rm d}\ln B_{\rm int}(0)/{\rm d} p\equiv {\rm d}\ln m_{\rm Ni}/{\rm d} p \simeq -0.3$~\%/GPa, see Fig.~\ref{fig:High-pressure}~{\bf c}.

The resistivity ($\rho$) under pressure experiments were preformed at pressures $p=0.0$, 0.61, 0.98, 1.49 and 1.97~GPa, Fig.~\ref{fig:High-pressure_resisitivity}. Following Refs.~\onlinecite{Wang_Arxiv_2023, Wu_PRB_2001, Liu_ScChinaMechAstr_2023, Fukamachi_JPSJ_2001, Kakoi_Arxiv_2023, Zhang_arxiv_2023, Hosoya_JPCS_2008, Wu_PRB_2023} the local minima in $\rho(T)$ could be associated with DW transition of unknown origin. The density wave transition temperature $T_{\rm DW}$ was obtained from parabolic fits of $\rho(T)$ in the vicinity of the local minima and it is plotted at Fig.~\ref{fig:High-pressure_resisitivity}~{\bf b}. $T_{\rm DW}$ decreases with increasing pressure as: $T_{\rm DW}(p)=132.3(1.6)-p\cdot25.7(1.2)$.

Our pressure data suggest that the DW order of unknown origin detected in resistivity experiments differs from the SDW magnetic order probed by means of $\mu$SR. Indeed, (i) the ambient pressure value of DW transition temperature [$T_{\rm DW}(p=0)\simeq 131$~K] is approximately 20~K lower compared to the SDW one [$T_{\rm SDW}(p=0)\simeq151$~K], and (ii) the SDW and DW ordering temperatures have opposite pressure behaviour. $T_{\rm SDW}$ increases with increasing pressure by approaching $T_{\rm N}\simeq 155$~K at 2.31~GPa, while $T_{\rm DW}$ decreases to $T_{\rm DW}\simeq 82$~K at $p=1.97$~GPa. Pressure, therefore, enhances the split between SDW and DW transitions from 20~K at $p=0$ to 70~K at $p\sim 2.0$~GPa.

\section{Discussion}

The discovery of superconductivity in Nd$_{0.8}$Sr$_{0.2}$NiO$_{2}$ films\cite{Li_Nature_2019} and in bulk La$_{3}$Ni$_{2}$O$_{7-\delta}$ under pressure\cite{Sun_Nature_2023} has sparked significant research interest in unraveling the pairing mechanism in nickel oxide systems. This discovery raises a crucial question: Are the mechanisms governing superconductivity in nickelates analogous to those observed in copper oxide superconductors?

In copper oxide superconductors, the phase diagram includes spin and charge order alongside superconductivity, and it is widely acknowledged that static and dynamic spin and charge orders play pivotal roles in the superconductivity mechanism.\cite{Tranquada_Nature_1995, Ghiringhelli_Sciense_2012, Chang_NatPhys_2012, Fradkin_RMP_2015, Keimer_Nature_2015, Kivelson_RMP_2003, Guguchia_PRL_2020} Elucidating competing orders in various nickel oxides is crucial for a comprehensive understanding of their electronic properties. In a similar vein, prior studies have revealed the presence of two distinct types of order in non-superconducting nickelate materials like La$_{2-x}$Sr$_x$NiO$_4$,\cite{Ricci_PRL_2021} La$_4$Ni$_3$O$_{10}$,\cite{Zhang_NatCom_2020} and hole-doped La$_4$Ni$_3$O$_8$.\cite{Zhang_PNAS_2016, Zhang_PRL_2019} Crucially, the ground state of La$_3$Ni$_2$O$_{7-\delta}$ also displays characteristics associated with a DW-like order.\cite{Wang_Arxiv_2023, Wu_PRB_2001, Liu_ScChinaMechAstr_2023, Fukamachi_JPSJ_2001, Kakoi_Arxiv_2023, Chen_arxiv_2023} This intriguing scenario is particularly noteworthy because these conditions align with those conducive to the emergence of high-temperature superconductivity in this material. This is evidenced by the suppression of a density wave like anomaly in resistivity as the material approaches the superconducting dome under pressure.\cite{Wang_Arxiv_2023}

Our experiments unveil the presence of commensurate magnetic order, as well as the critical temperature associated with the DW anomaly. One of the pivotal findings in our paper is the observation that pressure induces an elevation in the magnetic ordering temperature, underscoring the resilience of magnetism under external pressure. Significantly, this behavior stands in stark contrast to the substantial suppression, specifically by nearly 50~K within the same pressure range, of the transport anomaly associated with the DW instability (see Fig.~\ref{fig:High-pressure_resisitivity}). The distinctive pressure-induced response in La$_{3}$Ni$_{2}$O$_{7-\delta}$, as well as the results of the dipole-field calculations imply the simultaneous existence of two separate orders in its ground state: magnetic and most likely charge density wave orders. Remarkably, our results point to a pressure-enhanced decoupling between these two distinct orders. In cuprates or hole-doped nickelates like La$_{2-x}$Sr$_{x}$NiO$_{4}$, the relationship between spin and charge orders is notably intricate, with both orders being strongly intertwined and exhibiting similar responses to external parameters. However, this intricate interplay differs in La$_{3}$Ni$_{2}$O$_{7-\delta}$, where the behavior of spin and charge orders deviates from this pattern. Here, the response to external parameters reveals a distinct and less correlated behavior between spin and charge orders compared to the observed intertwining in cuprates and hole-doped nickelates. This distinction underscores the unique electronic characteristics and order interplay in La$_{3}$Ni$_{2}$O$_{7-\delta}$, adding to the complexity of the superconducting mechanism in this nickel oxide system.
\\

\section*{ACKNOWLEDGEMENTS}
Z.G. acknowledges support from the Swiss National Science Foundation (SNSF) through SNSF Starting Grant (No. TMSGI2${\_}$211750). The work of S.B., F.L., and I.M.E. is supported by the German Research
Foundation within the bilateral NSFC-DFG Project ER 463/14-1.

\section*{AUTHORS CONTRIBUTIONS}
R.K. conceived and supervised the project. D.J.G., I.P., and L.P.S. synthesized the sample and conducted x-ray characterization. R.K. performed the $\mu$SR experiments and analyzed the $\mu$SR data. T.J.H. calculated the muon stopping sites and dipole-field distributions for various possible magnetic structures. S.B., F.L., and I.M.E. calculated the spin susceptibility. V.S. and Z.G. conducted electrical transport experiments under pressure with contribution from M.B. R.K., T.J.H., and Z.G. wrote the manuscript with contributions from D.J.G., L.P.S., S.B., F.L., I.M.E., and H.L.

\section*{COMPETING INTERESTS}
The authors declare no competing interests.

\section*{INCLUSION \& ETHICS}
We have read the Nature Portfolio Authorship Policy and confirm that this manuscript complies the policy information about authorship: inclusion \& ethics in global research.



\section*{METHODS}

\subsection{Sample Preparation}
The La$_{3}$Ni$_{2}$O$_{7-\delta}$ sample was synthesized by means of the solid-state reaction as described in the Supplementary Note 1. The x-ray diffraction studies confirm the presence of a main fraction La$_{3}$Ni$_{2}$O$_{7-\delta}$ ($\delta\simeq 0$) and a small amount (of the order of 8\%) of lanthanum silica apatite impurity phase (Supplementary Note 1 and Supplementary Fig.~1).

\subsection{$\mu$SR experiments}

Zero-field (ZF) and weak transverse-field (WTF) $\mu$SR experiments were carried out at the Paul Scherrer Institute PSI, Switzerland. Experiments under ambient pressure were performed at the $\pi$E3 beamline using the GPS spectrometer.\cite{Amato_RSI_2017} Experiments under the quasi-hydrostatic pressure conditions were conducted at the $\mu$E1 beamline using the GPD spectrometer.\cite{Khasanov_HPR_2016,Khasanov_JAP_2022} A pressure up to 2.3~GPa was generated in a double-walled clamp type cell made of nonmagnetic MP35N alloy.\cite{Khasanov_HPR_2016} As a pressure transmitting medium, Daphne 7373 oil was used.

\subsection{Muon stopping sites and dipole-field calculations}

The muon stopping sites in La$_3$Ni$_2$O$_7$ were calculated by means of a DFT$+\mu$ approach\cite{blundell2023dft} using the MuFinder application.\cite{huddart2022mufinder} The dipole-field calculations utilised the \textsc{muesr} code.\cite{bonfa2018introduction}

\subsection{ZF-$\mu$SR data analysis procedure}

The fit to the experimental data was performed using the following functional form:
\begin{equation}
A(t)= A_{\rm s} [f_{\rm m}P_{\rm s,m}(t)+(1-f_{\rm m})P_{\rm s,nm}  ]+A_{\rm bg}P_{\rm bg}(t).
 \label{eq:asymmetry}
\end{equation}
Here, $A_{\rm s}/A_{\rm bg}$ and $P_{\rm s}(t)/P_{\rm bg}(t)$ are the initial asymmetry and the time evolution of the muon-spin polarization of the sample(s)/background(bg) contribution. The sample part is further divided into the magnetic (m) and nonmagnetic (nm) contributions with weights $f_{\rm m}$ and $1-f_{\rm m}$, respectively, and
\begin{equation}
P_{\rm s,m}(t) = \frac{2}{3}  \sum_i f_i e^{-\lambda_{{\rm T},i} t}\cos(\gamma_\mu B_{{\rm int},i}t) +
 \frac{1}{3} e^{-\lambda_{\rm L}}.
 \label{eq:commensurate}
\end{equation}
Here, $\gamma_\mu=851.616$~MHz/T is the muon gyromagnetic ratio, $\lambda$'s are the exponential relaxation rates, and $f_i$ is the volume fraction of $i-$th magnetic component. The indices 2/3 and 1/3 account for powder averaging, where 2/3 of the muon-spins precess in internal fields perpendicular (transversal, T) to the field directions and 1/3 remain parallel (longitudinal, L) to $B_{{\rm int},i}$.\cite{Schenck_book_1985, Schenk_book_1995, Smilga_book_1994, Karlsson_book_1995, Lee_book_1999, Yaouanc_book_2011, Blundell_book_2022}

\subsection{Resistivity experiments}

Experiments under ambient pressure were performed by using the Resistivity option hardware and software of the Quantum Design Physical Property Measurement System (PPMS). Experiments under pressure were performed using the same PPMS instrument by using Almax Easylab Pcell 15/30 module.\cite{Easylab}

\section*{DATA AVAILABILITY}
All relevant data are available from the authors. The data can also be found at the following link http://musruser.psi.ch/cgi-bin/SearchDB.cgi.

\end{document}